\documentclass[aps,prd,groupedaddress,superscriptaddress, twocolumn]{revtex4-2}

\pdfoutput=1

\usepackage{mathtools}
\usepackage{comment}
\usepackage{hyperref}
\usepackage{makecell}
\usepackage{graphicx} % Required for inserting images
\usepackage[caption=false]{subfig}
\usepackage{color}

\usepackage[normalem]{ulem}

\newsavebox{\measurebox}

\begin{document}
%\title{\boldmath Impact of  extreme ultraviolet radiation on liquid argon scintillation as a function of the xenon doping}
\title{Impact of extreme ultraviolet radiation on the scintillation of pure and xenon-doped liquid argon}

\newcommand{\gssi}{Gran Sasso Science Institute, L'Aquila 67100, Italy}
\newcommand{\icb}{ICB-UMR 6303 CNRS, Bourgogne  University, Dijon, France}
\newcommand{\apc}{APC, Universit\'e de Paris Cit\'e, CNRS, Astroparticule et Cosmologie, Paris F-75013, France}
\newcommand{\astrocent}{AstroCeNT, Nicolaus Copernicus Astronomical Center of the Polish Academy of Sciences, 00-614 Warsaw, Poland}
\newcommand{\princeton}{Physics Department, Princeton University, Princeton, NJ 08544, USA}
\newcommand{\ctp}{Mines Paris, PSL University, Centre for Energy Environment Processes (CEEP), 77300 Fontainebleau, France}
\newcommand{\ucdavis}{Department of Physics, University of California, Davis, CA 95616, USA}
\newcommand{\lpnhe}{LPNHE, CNRS/IN2P3, Sorbonne Universit\'e, Universit\'e Paris Diderot, Paris 75252, France}
\newcommand{\ihep}{Institute of High Energy Physics, Chinese Academy of Sciences, Beijing 100049, China}
\newcommand{\casb}{University of Chinese Academy of Sciences, Beijing 100049, China}
\newcommand{\williams}{Williams College, Physics Department, Williamstown, MA 01267 USA}
\newcommand{\cenpa}{Center for Experimental Nuclear Physics and Astrophysics, and Department of Physics, University of Washington, Seattle, WA 98195, USA}

\author{P.~Agnes}\affiliation{\gssi}
\author{Q.~Berger}\affiliation{\icb}
\author{M.~Bomben}\affiliation{\apc}
\author{M.~Campestrini}\affiliation{\ctp}
\author{M.~Caravati}\affiliation{\gssi}
\author{A.~F.~V.~Cortez}\affiliation{\astrocent}
%\author[\Napc,1]{D.~Franco\note{Corresponding Author}}\emailAdd{davide.franco@apc.in2p3.fr}
\author{D.~Franco}\email{davide.franco@apc.in2p3.fr}\affiliation{\apc}%\emailAdd{davide.franco@apc.in2p3.fr}
\author{C.~Galbiati}\affiliation{\princeton}
\author{G.~K.~Giovanetti}\affiliation{\williams}
\author{T.~Hessel}\affiliation{\apc}
\author{C.~Hidalgo}\affiliation{\princeton}
\author{S.~Hoceini}\affiliation{\ctp}
\author{C.~Houriez}\affiliation{\ctp}
\author{A.~Jamil}\affiliation{\princeton}
\author{P.~Kunz\'e}\affiliation{\gssi}
\author{J.~Machts}\affiliation{\apc}
\author{E.~Nikoloudaki}\affiliation{\apc}
\author{D.~Pailot}\affiliation{\apc}
\author{E.~Pantic}\affiliation{\ucdavis}
\author{C.~Savarese}\affiliation{\cenpa}
\author{P.~Stringari}\affiliation{\ctp}
\author{A.~Sung}\affiliation{\princeton}
\author{L.~Scotto Lavina}\affiliation{\lpnhe}
\author{J-M~Simon}\affiliation{\icb}
\author{H.~Vieira~de~Souza}\affiliation{\apc}
\author{M.~Wada}\affiliation{\astrocent}
\author{Y.~Wang}\affiliation{\ihep}\affiliation{\casb}
\author{Y.~Zhang}\affiliation{\ihep}

\collaboration{The X-ArT Collaboration}\noaffiliation

\begin{abstract}

The Xenon-Argon Technology (X-ArT) collaboration presents a study on the dynamics of pure and xenon-doped liquid argon (LAr) scintillation. Using two types of silicon photomultipliers  sensitive to different wavelength ranges,  we provide evidence in favor of a contribution from long-lived ($>$10~$\mu$s) extreme ultraviolet (EUV) lines emitted from argon atomic states, which enhances the light yield.   This component is present in both pure and xenon-doped LAr, becoming more pronounced at higher xenon concentrations, where it complements the traditional collisional energy transfer process. To explain this mechanism, we develop a comprehensive model of the Xe-doped LAr scintillation process that integrates both collisional and radiative contributions. Additionally, we investigate how xenon doping affects LAr scintillation light yield and pulse shape discrimination. Finally, we  hypothesize that the EUV component may explain the emission of spurious electrons, a known challenge in light dark matter searches using noble liquids. 

By characterizing the scintillation dynamics in Xe-doped LAr, identifying the long-lived EUV component, and exploring the potential origin of spurious electrons, this work lays the groundwork for optimizing detector performance and advancing the design and sensitivity of future noble liquid particle detectors. 

\end{abstract}

\maketitle

%%%%%%%%%%%%%%%%%%%%%%%%%%%%%%%%%%%%%%%%%%%%%%%%%%%%%%%%%%%%%%%%%%
\section{Introduction}
\label{sec:intro}
%%%%%%%%%%%%%%%%%%%%%%%%%%%%%%%%%%%%%%%%%%%%%%%%%%%%%%%%%%%%%%%%%%

Liquid argon (LAr) has emerged as a pivotal technology in both neutrino physics and dark matter searches because of its scalability, radiopurity, and ability to achieve high energy and spatial resolutions. The development of LAr sourced from deep underground, naturally depleted in cosmogenic $^{39}$Ar contamination \cite{DarkSide:2018kuk}, has further enhanced its appeal, positioning LAr as a key component in next-generation astroparticle experiments such as DarkSide-20k and ARGO~\cite{Franco:2015pha, DarkSide20k:2020ymr}. These massive LAr detectors are set to play a crucial role in advancing our understanding of neutrinos and exploring the frontiers of dark matter detection.

The capabilities of LAr can be extended by doping it with xenon, a technique that modifies the LAr response properties by making the scintillation process faster, increasing photon and ionization yields, and enhancing the photon attenuation length. This improvement results from the shift in the scintillation wavelength from 128 nm to 175 nm, which falls within the absorption band of VUV photosensors, enabling detection efficiencies as high as 30\%~\cite{Vogl:2021rba,Galbiati:2020eup}.   These features are particularly advantageous for large experiments designed for rare-event searches, which may benefit from lower energy thresholds, better topological reconstruction of event interactions, and higher detection efficiency. 

\begin{figure*}[t]
\centering
\sbox{\measurebox}{%
  \begin{minipage}[b]{.54\textwidth}
  \subfloat
    {\label{fig:figA}\includegraphics[width=1.05\textwidth]{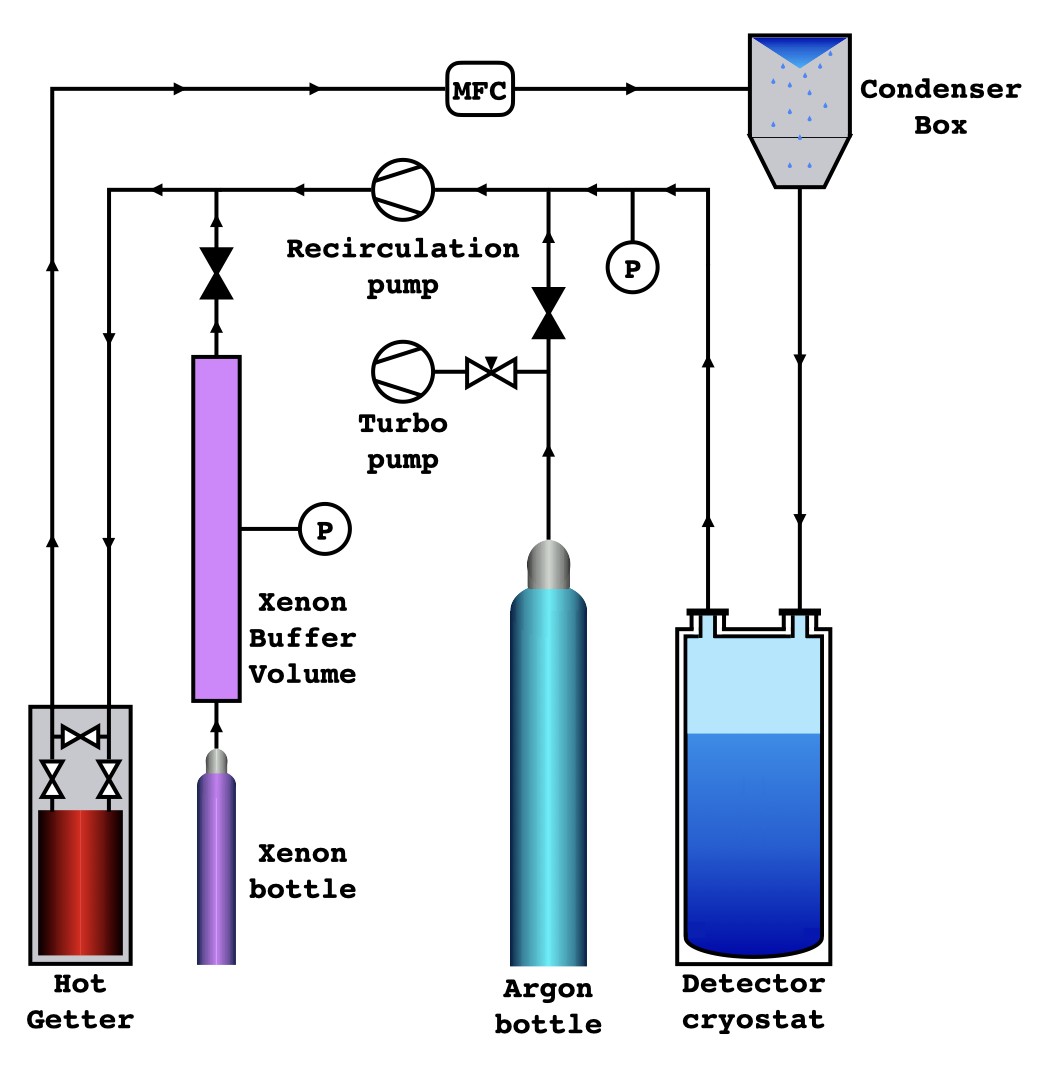}}%,height=7cm]
  \end{minipage}}
\usebox{\measurebox}\qquad
\begin{minipage}[b][\ht\measurebox][s]{.3\textwidth}
    \centering
    \subfloat
    {\label{fig:figB}\includegraphics[width=0.8\textwidth]{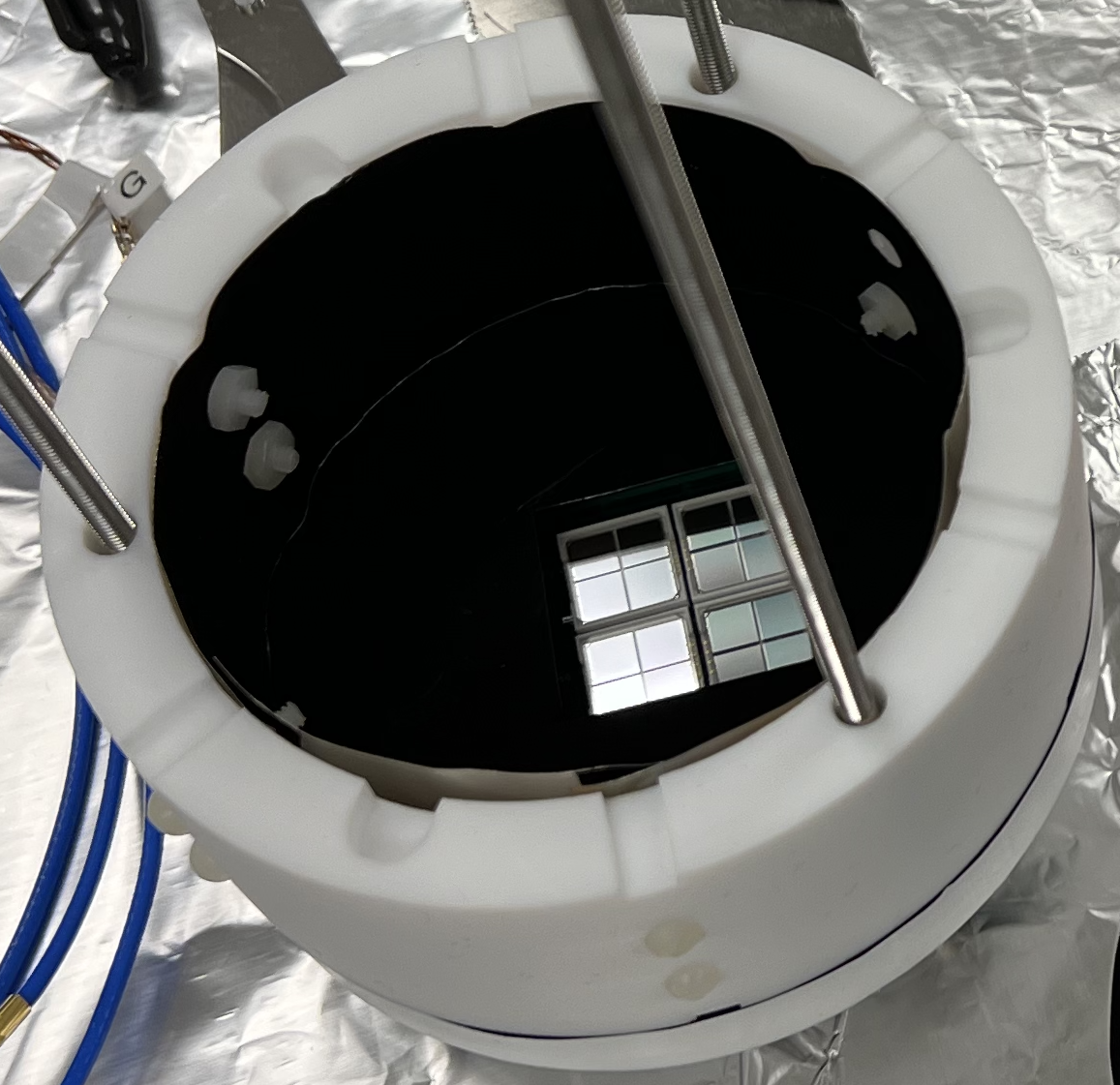}}%,height=2cm]
    \vfill
    \subfloat
    {\label{fig:figC}\includegraphics[width=0.8\textwidth]{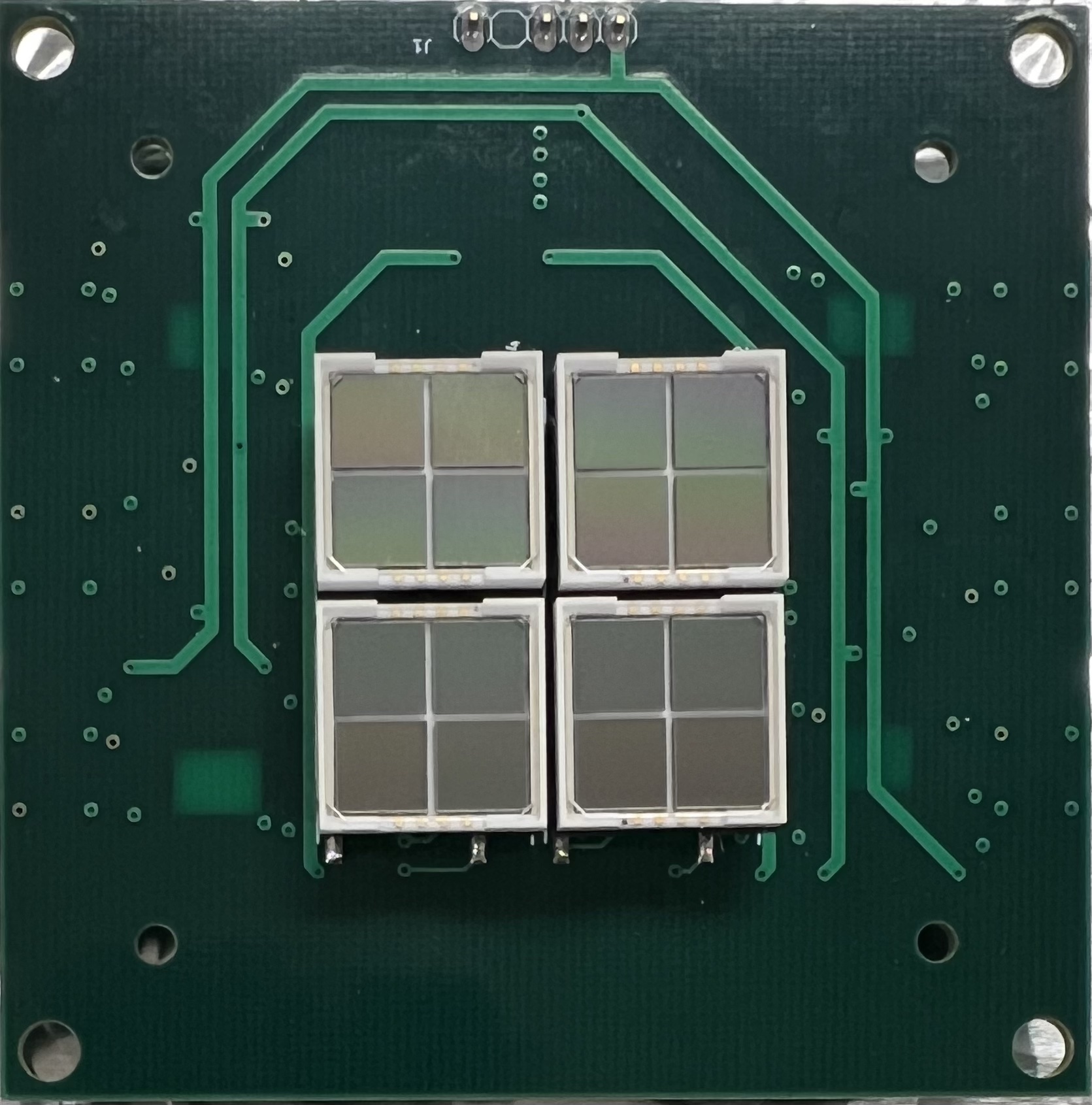}}%,height=2cm
\end{minipage}
\caption{{\bf Left:} schematic representation of the cryogenic setup for liquefaction, mixing, and purification of argon and xenon. The circles marked with ``P" represent pressure transducers, while the box marked as ``MFC" represents the mass flow controller. {\bf Top Right:} open PTFE detector chamber, lined with Lambertian Black\texttrademark\ foil, showing the bottom array of Hamamatsu SiPM modules. {\bf Bottom Right:} four Hamamatsu modules of VUV-sensitive SiPMs mounted on a custom-designed readout board.}
\label{fig:setup}
\end{figure*}

In this context, Xe-doped LAr emerges as an ideal medium to enhance sensitivity across a variety of physics cases, such as the detection of CNO, \textit{pep}, $^7$Be, and $^8$B neutrinos from the Sun via elastic scattering off electrons~\cite{Franco:2015pha}.  Additionally, dissolving $^{136}$Xe in LAr offers a promising avenue for searching for neutrinoless double beta decay~\cite{PhysRevD.106.092002}. Xe-doped LAr also holds potential for medical applications, such as Positron Emission Tomography \cite{Zabihi2024}, and for detecting core-collapse supernova neutrinos via coherent scattering off nuclei \cite{DarkSide20k:2020ymr}. Furthermore, its high ionization yield makes it well suited for exploring interactions of light dark matter particles, including GeV/c$^2$-mass WIMPs, galactic axions, dark photons, and sterile neutrinos~\cite{DarkSide-20k:2024yfq}.

%These changes may translate into lower energy thresholds, improved topological reconstruction of event interactions, and higher detection efficiency, beneficial for large experiments designed for rare-event searches. More specifically, these enhanced properties make Xe-doped LAr an ideal candidate for a wide range of applications like solar  physics by detecting CNO, \textit{pep}, $^7$Be, and $^8$B  neutrinos from the Sun through elastic scattering off electrons \cite{Franco:2015pha}. Additionally, dissolving $^{136}$Xe in LAr offers a promising avenue for searching for neutrinoless double beta decay~\cite{PhysRevD.106.092002}. Xe-doped LAr also holds potential for medical applications, such as Positron Emission Tomography \cite{Zabihi2024}, and for detecting core-collapse supernova neutrinos via coherent scattering off nuclei \cite{DarkSide20k:2020ymr}. Furthermore, its high ionization yield makes it well suited for exploring interactions of light dark matter particles, including GeV/c$^2$-mass WIMPs, galactic axions, dark photons, and sterile neutrinos~\cite{DarkSide-20k:2024yfq}.

This work, carried out within the X-ArT R\&D project, investigates the scintillation dynamics of Xe-Ar mixtures as a function of xenon concentration. For the first time,  we provide evidence for the presence of a radiative component induced by Extreme Ultra Violet (EUV) photons, which increases both light and ionization yields. This radiative mechanism complements the traditional collisional energy transfer process associated with xenon doping. By exploring these dynamics, we aim to optimize the scintillation properties of Xe-doped LAr and shed light on how radiative processes can affect both light yield and electron production in these systems. This understanding is crucial for advancing the design of next-generation detectors for neutrino physics, dark matter detection, and other applications.

Finally, our results also suggest that the EUV-induced component could play a role in the emission of spurious electrons in dual-phase TPC,  a well-known issue in  experiments searching for light dark matter using LAr or liquid xenon (LXe) targets \cite{DarkSide-20k:2024yfq, XENON:2021qze}.

\section{The experimental setup}
\label{sec:setup}

The detector consists of a single-phase liquid argon (LAr) chamber designed for studying xenon-doped liquid argon. As shown in figure~\ref{fig:setup}, the geometry is cylindrical  with inner height and diameter of 4.5 cm and 9.5 cm respectively. The chamber structure is made of polytetrafluoroethylene (PTFE), and its inner surfaces are lined with Acktar Lambertian Black\texttrademark\ foil to minimize photon reflections. The chamber is equipped with fourth-generation VUV-sensitive silicon photomultipliers (SiPMs) Hamamatsu S13371, for detecting scintillation light. The 32 photosensors are packaged in eight modules, each containing four units, and read out as a single channel. Four modules are arranged on each of the two readout boards, mounted at the chamber's top and bottom. Four out of the eight SiPM modules are equipped with quartz windows (WW) with a cutoff in transmission around 155 nm, making them primarily sensitive to Xe$_2^*$ photons, which are emitted with a wavelength of 175 nm. The remaining SiPMs are windowless (WL), allowing them to detect a broader range of wavelengths, including 128-nm photons from Ar$_2^*$ de-excitation.  Optical fibers connected to a Hamamatsu 402 nm picosecond laser are used for calibrating the SiPMs, and PT100 temperature sensors are installed to ensure precise monitoring of thermal conditions inside the chamber.

The test chamber, hosted in a double-walled vacuum jacketed dewar, is immersed in and filled with the Ar-Xe mixtures under study. The Xe-doped argon is kept liquid and free of contamination by a custom-designed cryogenic setup. This system, sketeched in figure~\ref{fig:setup}, consists of a bespoke pump that extracts the gaseous mixture from the top of the dewar flask and forces it through a SAES Micro-Torr hot getter, capable of purifying the stream to sub-ppb level of contaminants. The clean Ar-Xe mixture is then liquified in a condenser box, driven by a 90~W Cryomech pulse tube refrigerator, before returning into the dewar. The system is continuously monitored with a mass flow controller, pressure meters, and temperature sensors, and it includes passive safety mechanisms such as a rupture disk to prevent over-pressurization. 
The xenon injection system, integrated within the cryogenic setup, is designed to introduce controlled amounts of xenon into the LAr chamber. It comprises of a xenon gas source connected to a buffer volume, temporarily holding the gas before injection. A valve controls the flow of xenon into the argon stream, and a pressure meter monitors the pressure in the buffer.  This configuration allows for a control of the amount of xenon injected into the system at about 20\% accuracy, enabling careful monitoring and adjustment of gas concentrations to achieve the desired mixture. The primary sources of uncertainty are from the buffer volume and the total argon mass, each contributing about 10\%. Additional  sources of uncertainty, such as pressure fluctuations and the tick size of the pressure transducer, are sub-dominant.

%%%%%%%%%%%%%%%%%%%%%%%%%%%%%%%%%%%%%%%%%%%%%%%%%%%%%%%%%%%%%%%%%%
\section{The EUV contribution to liquid argon scintillation }
\label{sec:lar}
%%%%%%%%%%%%%%%%%%%%%%%%%%%%%%%%%%%%%%%%%%%%%%%%%%%%%%%%%%%%%%%%%%

\begin{figure}
  \includegraphics[width=0.48\textwidth]{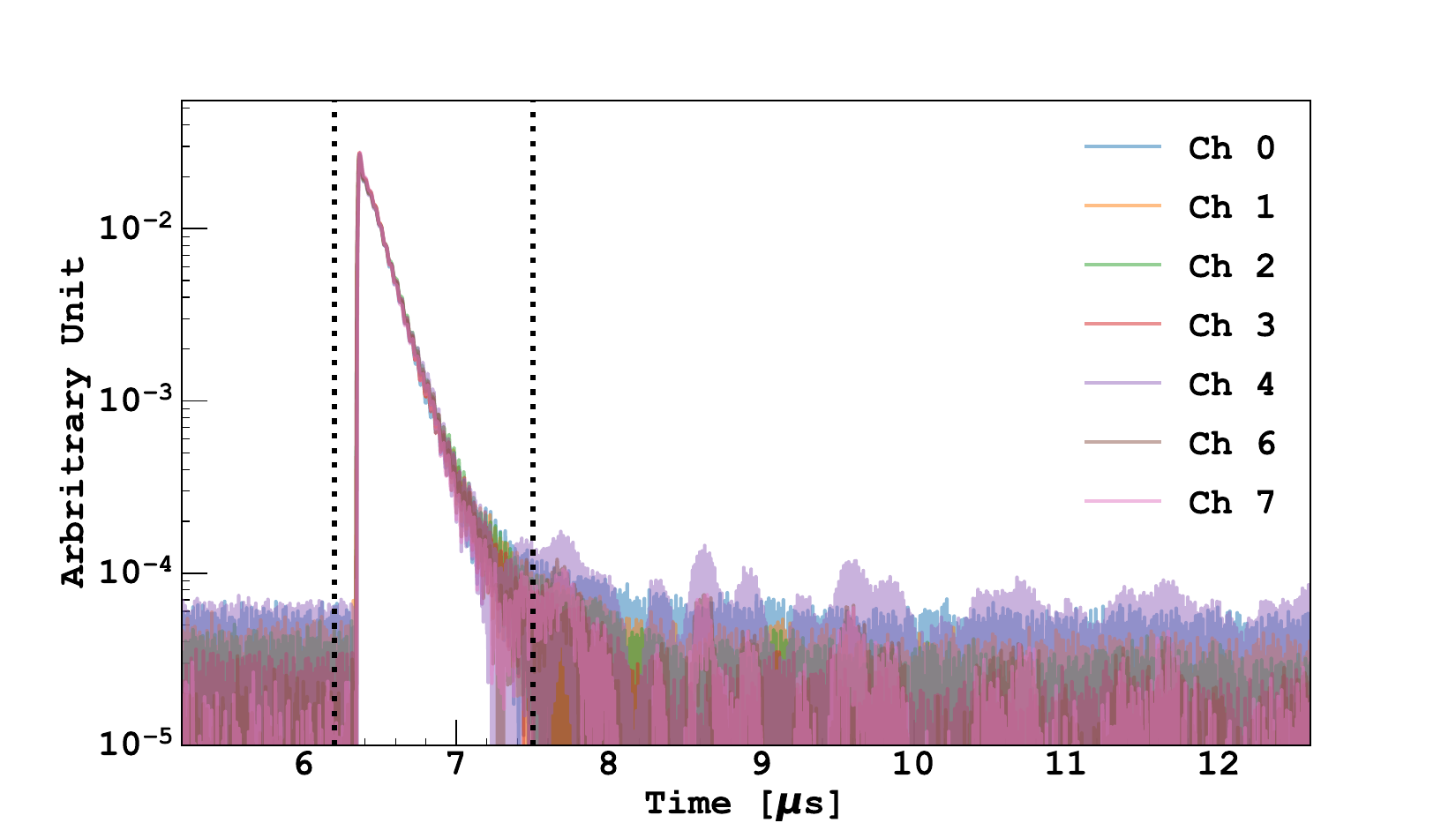}
  \includegraphics[width=0.48\textwidth]{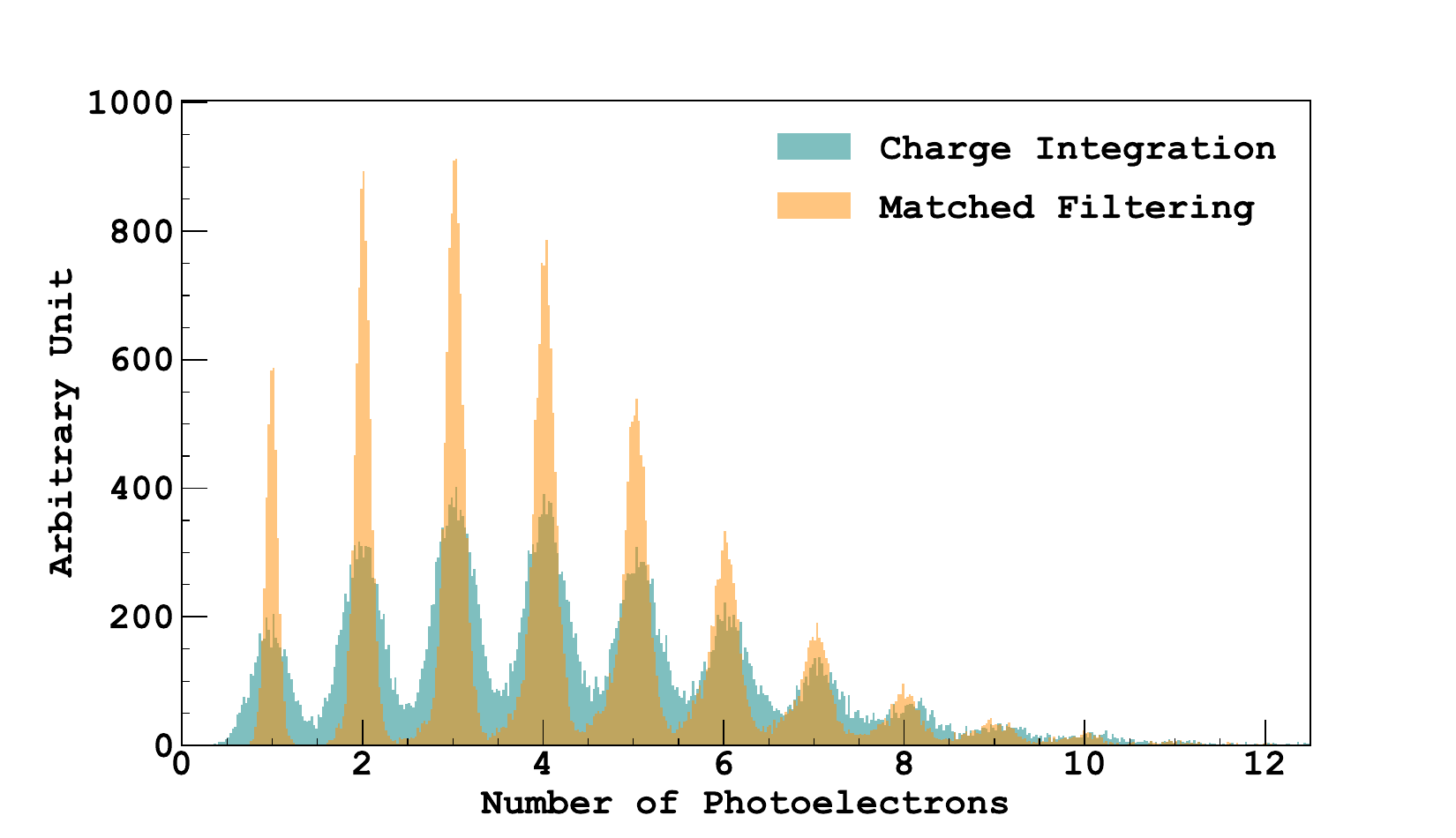}
  \caption{Left. SiPM time responses to laser photons. The dashed vertical lines in the left plot represent the range used for extracting the response template to convolve the scintillation response. Right. SiPM charge response obtained through waveform integration (teal) and matched filter analysis (orange), using the  time-reversed SiPM response. } 
  \label{fig:spe}
\end{figure}

Characterizing the scintillation properties of Xe-doped LAr first requires measuring and analyzing the response of pure LAr in the setup to constrain the Ar$^{*}_{2}$ triplet lifetime and assess LAr purity. Averaged waveforms were generated from a calibration run with $10^5$ triggers, dominated by the activity of a $^{60}$Co $\gamma$-source placed next to the cryostat. We analyzed waveforms from WL-SiPMs only (as WW-SiPMs are not sensitive to 128-nm photons), selecting events within the 300-700 photoelectron (pe) range.

\begin{figure}
\includegraphics[width=0.49\textwidth]{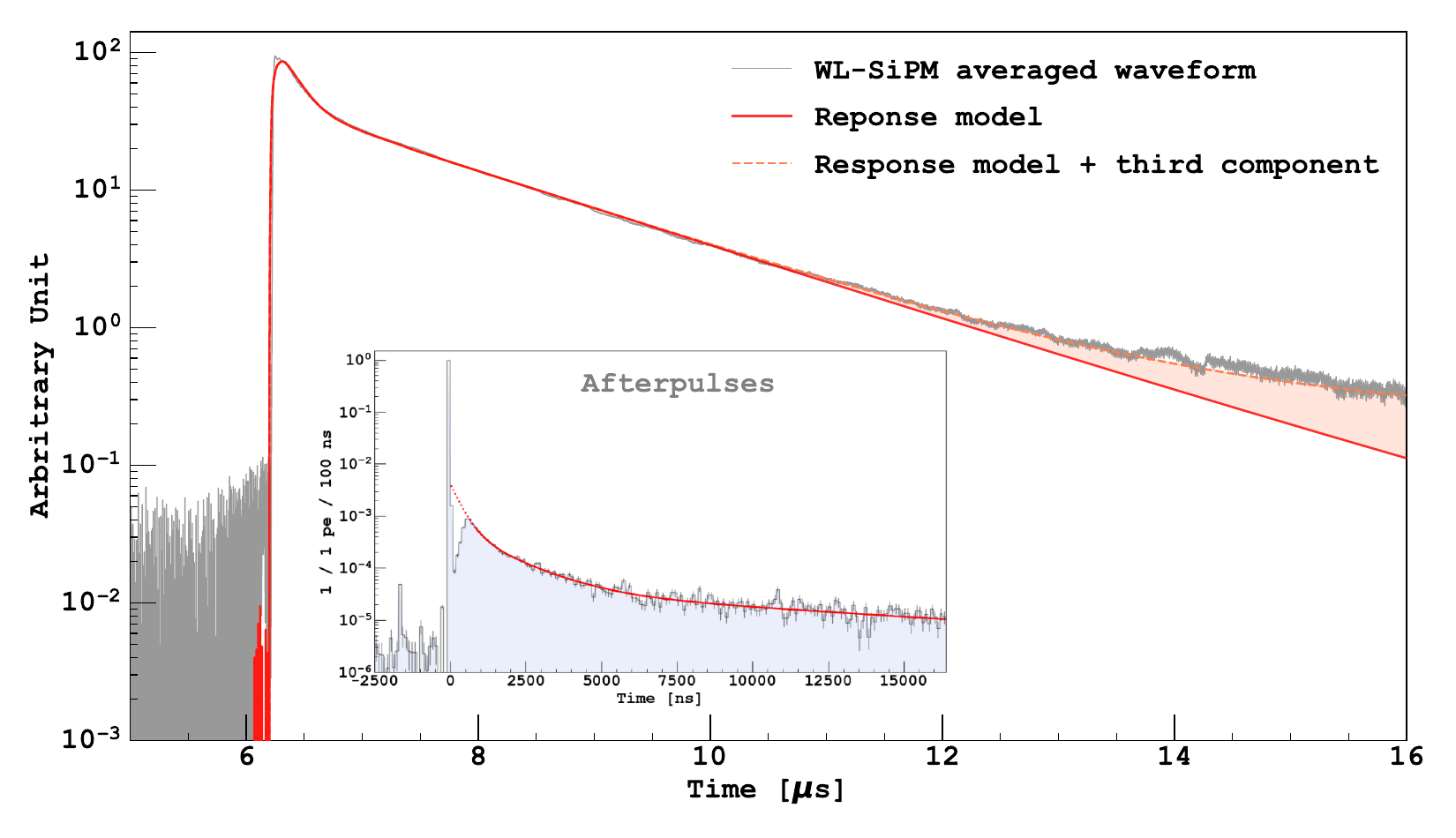}
\caption{Fit of the averaged waveform from pure LAr scintillation, as acquired by WL-SiPMs. The red line represents the response model from the convolution of the scintillation response and the SiPM response, including the contribution from APs, as measured in X-ArT and shown in the inset. The discrepancy between such a model and data (red shaded area) is recovered (dashed line) when including a third component. 
}
\label{fig:lar_fit}
\end{figure}

The LAr scintillation response is modeled as the sum of two exponentials, representing the singlet and triplet de-excitations. This model is convolved with the SiPM time response, shown in figure~\ref{fig:spe}, which was determined through laser calibration measurements. The resulting response is further convolved with the time profile of SiPM afterpulses (APs), characterized \textit{in situ} also exploiting laser calibrations in the single photoelectron regime.  The inset of figure~\ref{fig:lar_fit} shows the time response of APs following a laser signal,  equivalent to a single photoelectron to prevent the generation of multiple APs.  The AP distribution is normalized to the number of selected laser triggers.  The AP response is empirically modeled with three exponentials. The fit yields characteristic times of 80 $\pm$ 13 ns, 350 $\pm$ 30 ns, and 3010 $\pm$ 433 ns, along with associated emission probabilities of 26.8 $\pm$ 6.2\%, 5.3 $\pm$ 0.4\%, and 1.4 $\pm$ 0.4\%, respectively.

The overall response model accurately reproduces the acquired waveforms, as shown in figure~\ref{fig:lar_fit}, up to $\sim$5~$\mu$s from the start of the waveform, resulting in a LAr triplet lifetime equal to 1.49 $\pm \ 0.01~\mu$s. However, at longer times, the model fails to reproduce the averaged waveform, which exhibits an additional slow component. Indeed, the discrepancy between the data and the model is recovered by adding a third LAr decay component.  The fitted third component accounts for 11.6 $\pm$ 0.2\% of the Ar$_2^*$ contribution and has a lifetime that can be approximated as constant ($>10$~$\mu$s) when compared to the acquisition window.

Such a long tail  was also observed by DEAP-3600~\cite{DEAP:2020hms} using the tetraphenyl-butadiene (TPB)  wavelength shifter~\cite{Benson2018}.  In the literature, it has been suggested that this long component may be due to fluorescence from TPB or Teflon~\cite{Asaadi:2018ixs,Jerry:2010zj,NEXT:2022cmg}, a hypothesis ruled out by the current measurement due to the absence of TPB in the setup and the use of black foil covering the internal walls. Furthermore, we can exclude the possibility that this component is due to near-infrared photons, also reported in literature~\cite{Alexander:2016zpd},  as WW-SiPMs do not detect any significant contributions from photons with wavelengths greater than 155 nm.

To account for this additional component, we hypothesize that emission in the EUV range contributes significantly, providing a viable explanation for the observed long tail. In fact,    EUV range photons~\cite{Gehman:2011xm} are  emitted primarily through direct de-excitation of 
neutral argon (Ar$^{*}$) and singly or multiple ionized argon (Ar$^{q+}$)~\cite{JMAjello_1990,10.1063/1.4981535}.

The emission spectrum of Ar$_2^*$ dimers between 110 and 155 nm has been well characterized through spectroscopy~\cite{Neumeier:2015ori}, with no additional lines observed beyond the 128~nm emission line from singlet and triplet states. In this same range, the NIST database~\cite{NIST_ASD} reports a few lines around 130 nm with lifetimes of tens of seconds, but these originate from the formation of highly ionized argon states, such as Ar$^{12+}$ (or Ar XIII), which are rarely produced by radiogenic ionization particles.

Below this range, NIST  lists approximately 1,200 emission lines between 10 and 110 nm, with Ar-I displaying peaks in the 80–90 nm range, while Ar-II emission extends down to 40 nm. Nevertheless, photon emission at wavelengths below 78 nm is not expected to contribute to scintillation, as these photons can ionize argon itself, given its first ionization energy of $\sim$15.8 eV.

The long scintillation component observed could be explained by emission from metastable atomic argon states in the EUV range. Metastable states are widely reported in the literature, from the ultraviolet~\cite{PhysRev.95.892} to the infrared~\cite{Schef2004}, with lifetimes spanning from microseconds to seconds. Despite their characterization in the EUV remains more challenging due to experimental constraints, evidence of metastable states in singly ionized argon has been reported in ref.~\cite{Lundin2007}, with measured lifetimes of a few seconds and emissions in the EUV around 65 nm. These long-lived states suggest that EUV emission could persist over extended timescales, which aligns with the observed slow scintillation component and further supports the hypothesis of EUV emission contributing to the LAr scintillation dynamics.

%However, transition probabilities in this region are challenging to measure. Some states, such as the singly charged 3s$^2$3p$^4$ ($^1$D)3d $^2$G$_{7/2,9/2}$ states of Ar$^+$, are known to be metastable~\cite{Lundin2007}, emitting in the EUV (65 nm) with a  measured lifetime of a few seconds. 

In summary, the observed long tail in the  scintillation cannot be explained by known LAr VUV (or longer-wavelength) contributions, nor by instrumental effects such as afterpulses or fluorescence from passive materials in the detector. The EUV contribution from excited argon atoms and ions, well-established in the literature, provides a compelling explanation to the late photon emission. This hypothesis is further supported by the analysis of the Xe-doped LAr response, discussed in the next section.

%%%%%%%%%%%%%%%%%%%%%%%%%%%%%%%%%%%%%%%%%%%%%%%%%%%%%%%%%%%%%%%%%%
\section{The Xe-doped liquid argon scintillation}
\label{sec:arxe_mechanism}
%%%%%%%%%%%%%%%%%%%%%%%%%%%%%%%%%%%%%%%%%%%%%%%%%%%%%%%%%%%%%%%%%%

To investigate the scintillation mechanism in Xe-doped LAr, we analyzed waveforms from pairs of WW and WL channels. Both channels were triggered simultaneously to ensure data consistency. A WL-channel detects the full range of scintillation wavelengths, while a WW one is specifically sensitive to the 175 nm emission from Xe$_2^*$ excimers. By fitting the waveforms from both channels simultaneously, we can partially disentangle the contributions of different wavelength components and gain insights into how xenon doping affects the scintillation properties of liquid argon.

\begin{figure}
  \includegraphics[width=0.49\textwidth]{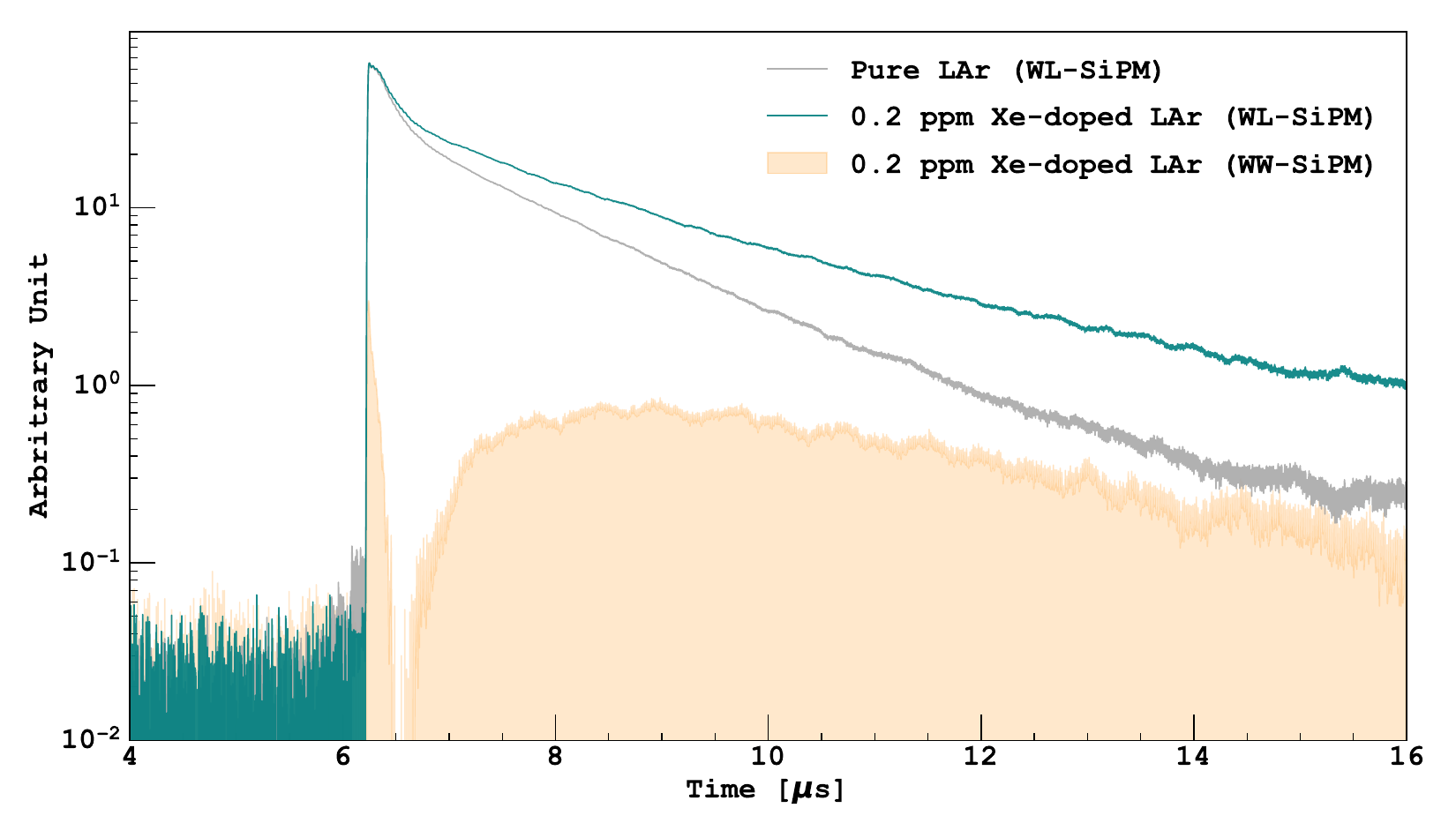}
  \caption{Comparison between the scintillation pulse shapes in pure liquid argon (grey) and with 0.2 ppm Xe doping (teal). The long pulse observed with the WL-SiPM and the low-amplitude signal from the WW-SiPM (orange shaded area) indicate the presence of the 150 nm component from the decay of the ArXe dimer ($\tau \sim 5~\mu$s)~\cite{Neumeier:2015ori}.}
  \label{fig:subppm}
\end{figure}
Doping LAr with just 0.2 ppm xenon introduces a slow component detectable with WL-SiPMs, as shown in figure~\ref{fig:subppm}, but not with WW-SiPMs (insensitive to wavelengths below 155 nm). This contribution arises from the de-excitation of the ArXe$^*$ dimer, with a lifetime of approximately 5~$\mu$s~\cite{Neumeier:2015ori} ($\tau_{150}$), produced via the reaction Ar$_2^*$ + Xe $\rightarrow$ ArXe$^*$ + Ar.

\begin{figure*}
  \centerline{\includegraphics[width=0.9\textwidth]{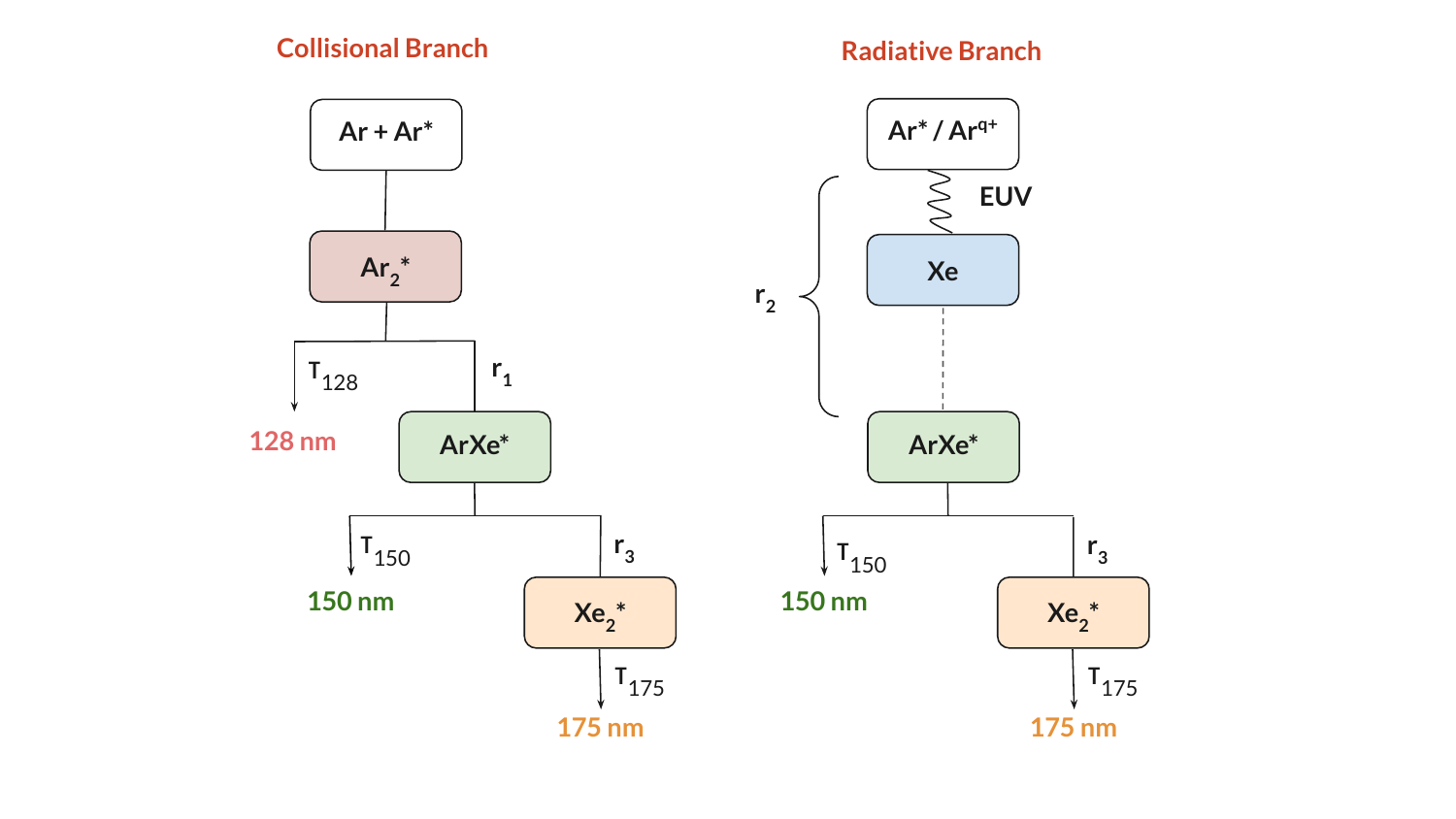}}
  \caption{Simplified schematic of the collisional and radiative processes underlying ArXe scintillation. Here, $r_i$ with $i = \{1, 2, 3\}$, and $\tau_i$, where $i = \{128, 150, 175\}$ nm, denote the process rates and decay times. The latter are associated with the de-excitations of Ar$_2^*$, ArXe$^*$, and Xe$_2^*$, respectively.}
  \label{fig:arxe_scheme}
\end{figure*}

As the xenon concentration increases, the reaction ArXe$^*$ + Xe $\rightarrow$ Ar + Xe$_2^*$ becomes increasingly probable. This process becomes dominant when its rate exceeds the inverse of the ArXe$^*$ dimer’s decay time ($1 / \tau_{150}$). This collisional component is illustrated schematically in figure~\ref{fig:arxe_scheme}.

More specifically, the  modeling of this branch involves two main components: (1) a direct LAr scintillation component, which occurs when the Ar$_2^*$ dimer is isolated and unable to transfer energy collisionally to a xenon atom; and (2) a component originating from the ArXe$^*$ dimer, which can either decay directly or transfer its energy to form a Xe$_2^*$ dimer. Both components are detectable by WL-SiPMs, but only the component originating from ArXe$^*$ is detectable by WW-SiPMs via the Xe$_2^*$ de-excitation.

The model has to take into account the competition between Ar$_2^*$ triplet decay, emitting 128-nm photons with $\mathcal{O}$(1~$\mu$s) characteristic  time,  

\begin{equation}
Ar_2^* \rightarrow Ar + Ar + \gamma_{128},\label{eq:ar1}
\end{equation}

\noindent and the formation of  an ArXe$^*$ excimer with rate $r_1$, 

\begin{equation}
Ar_2^* + Xe \rightarrow ArXe^* + Ar\label{eq:ar2} 
\end{equation}

The singlet state decays too rapidly  to contribute to ArXe$^*$ formation and  therefore it is not considered in this competition.

The ArXe$^*$, in turn, can either decay, emitting 150-nm light with decay constant $\tau_{150}$, or combine with a Xe atom to form a Xe$^*_2$ dimer with rate $r_3$:

\begin{eqnarray}
ArXe^* &\rightarrow& Ar + Xe + \gamma_{150}, \label{eq:arxe} \\
ArXe^* + Xe &\rightarrow& Xe_2^* + Ar. \label{eq:arxe2}
\end{eqnarray}

The Xe$_2^*$ de-excitation time scale is negligible compared to the characteristic times involved in the processes, and hence it is assumed to decay instantaneously.

The time evolution of the numbers of singlet  states is expressed by 
\begin{equation}
\frac{dN_{Ar,1}}{dt}(t) = - \frac{1}{\tau_s} \, N_{Ar,1}(t),     
\end{equation}

\noindent with $\tau_{s}$ the Ar$_2^*$ singlet  lifetime.  The time evolution of  triplet  states, with lifetime $\tau_{128}$,  is derived from eq.~\ref{eq:ar1} and eq.~\ref{eq:ar2} as

\begin{equation}
\frac{dN_{Ar,3}}{dt}(t) = - \frac{1}{\tau_{128}} \, N_{Ar,3}(t) - r_1 \, N_{Ar,3}(t).
\end{equation}

The competition between triplet decay and ArXe$^*$ formation is introduced through the $k_1$ rate, defined as 

\begin{equation}
k_1 = r_1 + \frac{1}{\tau_{128}}.
\end{equation}

The probability density function describing the time evolution of 128-nm photon emission is then modeled as

\begin{equation}
\label{eq:f128}
f_{128}(t) =   \frac{\alpha}{\tau_{s}} \, e^{-t/ \tau_{s}} + (1-\alpha) \, k_1 \, e^{-k_1 t},
\end{equation}

\noindent where $\alpha$ is the fraction of de-excitations from singlet states, which corresponds in pure LAr to the probability of populating singlet states.

\begin{figure*}
  \includegraphics[width=0.7\textwidth]{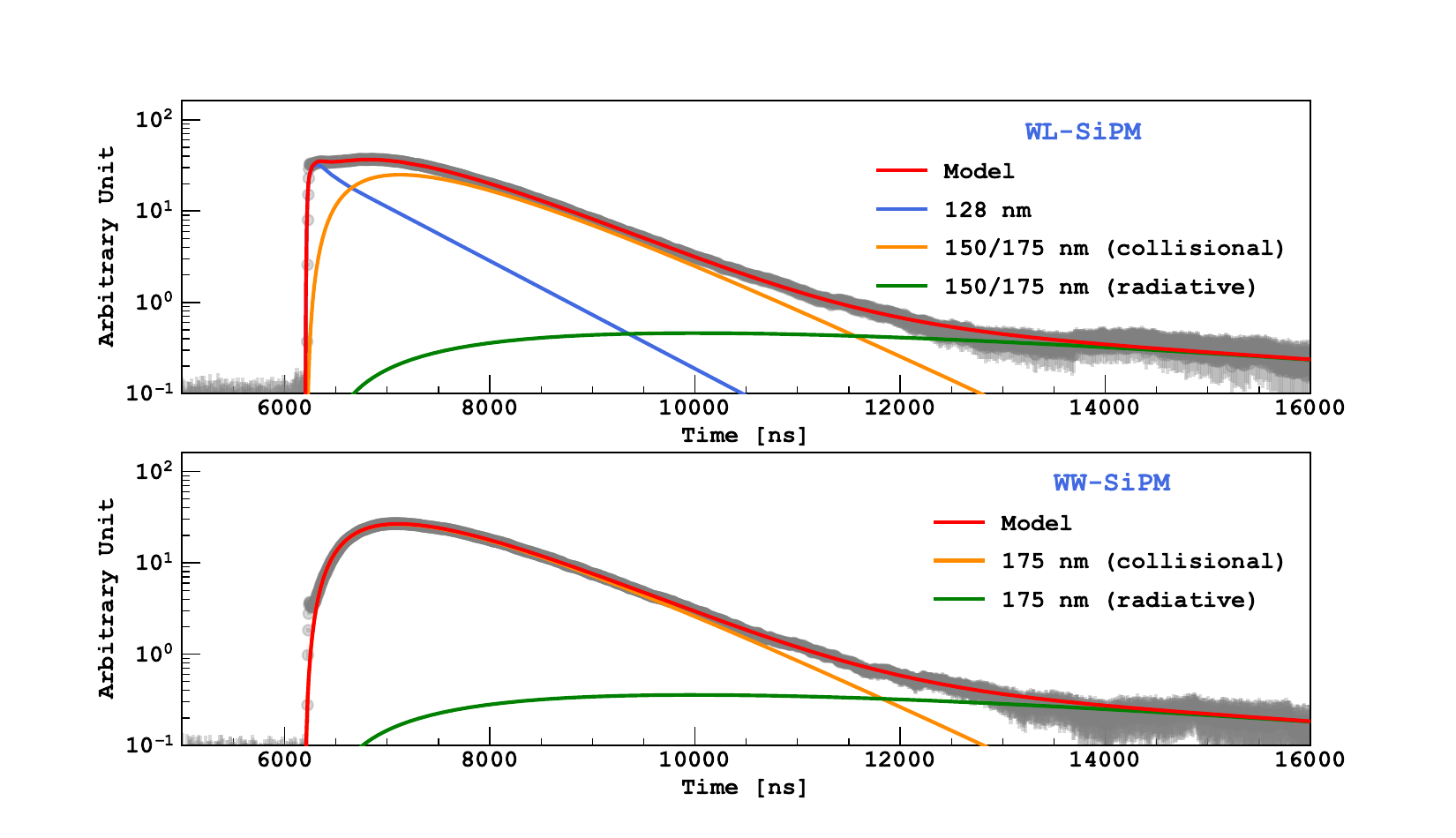}
  \caption{Simultaneous fit of averaged waveforms acquired with WL- and WW-SiPMs and 2.7-ppm Xe-doped LAr, using the model (red line) from eqs.~\ref{eq:ww} and \ref{eq:wl}. The contribution from direct Ar$^*_2$ de-excitation (blue) and from the collisional (yellow) and radiative (green)  branches are also shown.} 
  \label{fig:double_fit}
\end{figure*}

The evolution of the number of  excited ArXe dimers induced by collisional energy transfer (eq.~\ref{eq:arxe} and eq.~\ref{eq:arxe2}) is expressed by

\begin{equation}
\label{eq:diff_coll}
\frac{dN^c_{ArXe}}{dt}(t) = r_1 \, N_{Ar,3}(t) - r_3 \, N^c_{ArXe}(t) - \frac{1}{\tau_{150}} \, N^c_{ArXe}(t),
\end{equation}

\noindent which results in the probability density function, $f_{c}(t)$, of emitting 150- and 175-nm photons 
\begin{equation}
\label{eq:fc}
f_{c}(t) = \frac{k_1\, k_3}{k_1 - k_3} \, e^{-k_3\, t} \, (1 - e^{-(k_1-k_3) \, t}),
\end{equation}

\noindent where $k_3 = r_3 + \frac{1}{\tau_{150}}$ is the effective rate arising from the competition between ArXe$^*$ de-excitation and Xe$_2^*$ formation.

EUV photons, directly emitted by Ar$^*$ and Ar$^{q+}$ via metastable states, can participate in the formation of ArXe$^*$ dimers. EUV photons with wavelengths smaller than $\sim$100~nm (or energy greater than 12.1~eV) are sufficiently energetic to ionize xenon atoms, which may also explain the observed increase in ionization yield in Xe-doped LAr~\cite{KUBOTA1974393}. The resulting ionized xenon can then interact with argon atoms to form ArXe$^*$ dimers, which subsequently decay and contribute to the scintillation signal. This radiative contribution,  which represents the emission of multiple EUV lines from Ar$^*$ and Ar$^{q+}$ de-excitation to Xe$^+$ production and ArXe$^*$ formation, is effectively modeled here as a single de-excitation process with rate $r_2$, as illustrated in figure~\ref{fig:arxe_scheme}. At this stage, we lack the sensitivity to distinguish the contributions from the different EUV lines, which are expected to have much longer lifetimes than the acquisition window ($\sim$20~$\mu$s). As a result, we are unable to assess the time profile of the EUV emission. This will be the focus of a next X-ArT data campaign, with an acquisition window extended to the millisecond scale.

The probability density function of emitting 150- and 175-nm photons  as a result of the radiative excitation of ArXe is derived similarly to that of the collisional branch, and is expressed as 
\begin{equation}
\label{eq:fr}
f_{r}(t) = \frac{r_2\, k_3}{r_2 - k_3} \, e^{-k_3 t} \, (1 - e^{-(r_2-k_3) \, t}).
\end{equation}

The signal observed by WL-SiPMs is then derived from eqs.~\ref{eq:f128}, \ref{eq:fc}, and \ref{eq:fr} as the sum of contributions from Ar$^*_2$ de-excitations and the  collisional and radiative branches,

\begin{eqnarray}
\label{eq:wl}
f_{WL}(t) &=& \frac{A}{\varepsilon_{1}}  f_{128}(t) + \frac{B_1}{\varepsilon_{2}}  f_{c}(t) + \frac{C_1}{\varepsilon_{2}} \, f_{r}(t),
\end{eqnarray}

\noindent where $A$, $B_1$, and $C_1$ are the amplitudes associated to the direct Ar$_2^*$ de-excitation and to the collisional and radiative components, respectively, and  $\varepsilon_{1}$ and $\varepsilon_{2}$ are the SiPM photo-detection efficiencies (PDEs) at 128~nm (14.7$\substack{+1.1 \\ -2.4}$~\%~\cite{Pershing:2022eka}) and at 150-175~nm (20.5 $\pm$ 1.1~\%~\cite{Gallina:2022zjs}).

As WW-SiPMs can detect only wavelengths above 155 nm, they are sensitive to only Xe$_2^*$ de-excitation, and thus the signal is modeled as

\begin{eqnarray}
\label{eq:ww}
f_{WW}(t) &=&   \frac{B_2}{\varepsilon_{2}} \, f_{c}(t) + \frac{C_2}{\varepsilon_{2}} \, f_{r}(t).
\end{eqnarray}

\noindent with $B_2$ and $C_2$ the amplitudes of the collisional and radiative branches.  Both $f_{WL}(t)$ and $f_{WW}(t)$ are numerically convoluted with the SiPM response, as discussed in the previous section.

\begin{figure*}
\includegraphics[width=0.80\textwidth]{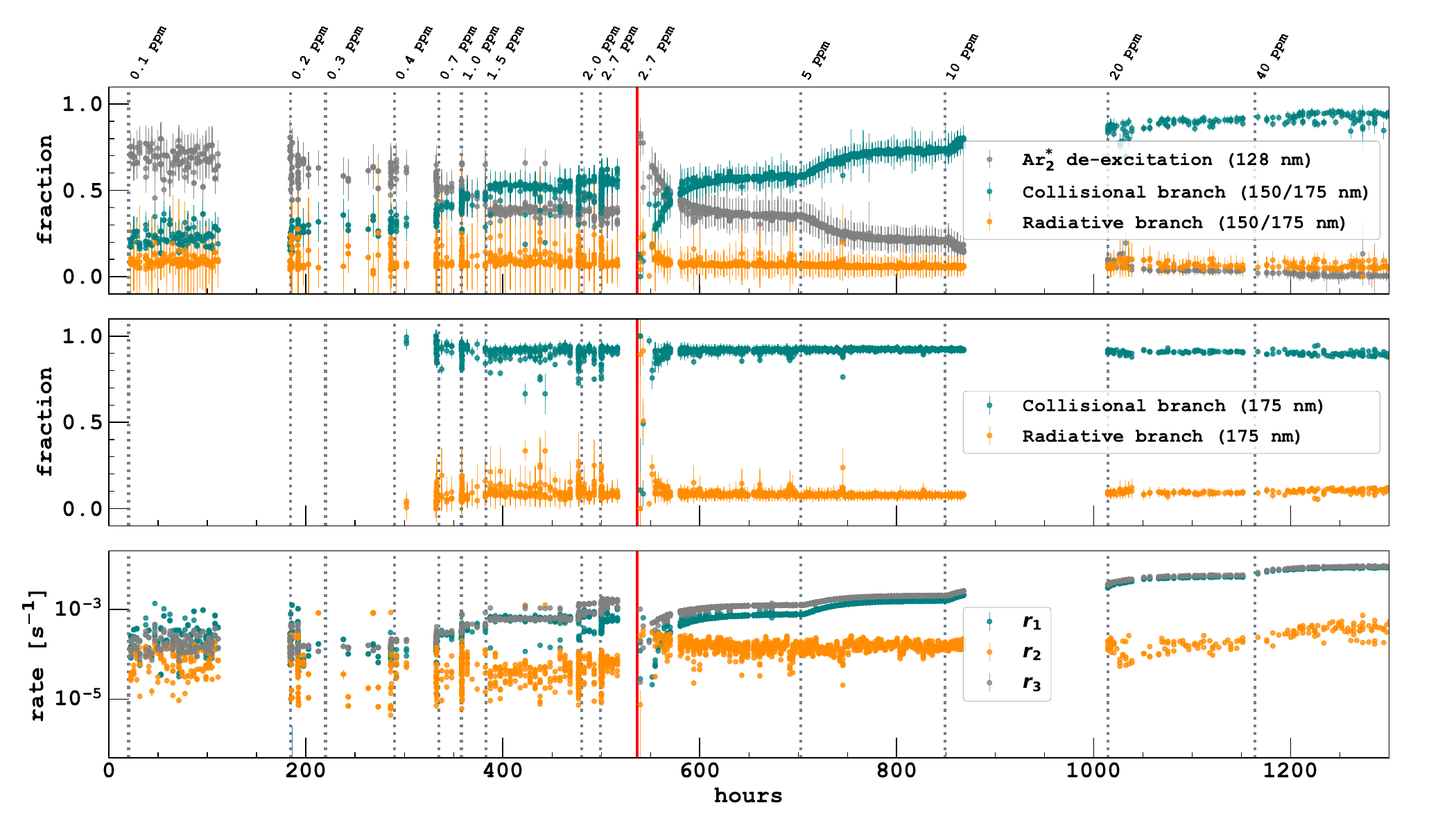}
  \caption{Evolution of the fractions of direct LAr scintillation, and collisional and radiative components for WL-SiPMs (top) and WW-SiPMs (middle) over time. The bottom plot shows the time evolution of the process rates $r_1$, $r_2$, and $r_3$. Dashed vertical lines indicate the times of xenon injections, while the red line marks the start of the second campaign of measurements from pure LAr.} 
  \label{fig:evolution}
\end{figure*}

Figure~\ref{fig:double_fit} shows averaged waveforms acquired with 2.7 ppm Xe-doped LAr exposed to a $^{60}$Co source.  WL and WW-SiPM waveforms are simultaneously fitted using eqs.~\ref{eq:wl} and \ref{eq:ww}, with $r_1$, $r_2$, and $r_3$ as common parameters, and $\tau_s$, $\tau_{128}$, and $\tau_{150}$ fixed to their nominal values. The fit  indicates a non-zero contribution from the radiative branch (green line in figure \ref{fig:double_fit}), as the collisional process (orange line) alone is insufficient to fully explain the observed data.  The agreement between the data and the model, incorporating the radiative branch, further supports the hypothesis of EUV contributions from the de-excitation of atomic argon.

This measurement is part of a broader data acquisition campaign aimed at studying the effects of xenon doping in LAr. Indeed,  we conducted two data acquisition campaigns by injecting Xe into LAr. Each campaign began with about  14 days of pure LAr circulating through a getter to achieve the desired purity. In the first set of measurements, the Xe concentration was gradually increased from 0.1 ppm to 2.7 ppm. In the second set, the concentration was increased from 2.7 ppm to 40 ppm. After each Xe injection, we exposed the chamber to radioactive sources ($^{60}$Co, $^{22}$Na, $^{241}$AmBe) and collected data over several hours. 

For each run, we analyzed the data by fitting pairs of averaged waveforms from the WW and WL channels, repeating the fit for all possible permutations of these two types (WW and WL) of channels. Given that the statistical errors are very small due to the high statistics, we used this approach to assess potential systematics. The result for each run is obtained by averaging the fit results, while the systematic error is evaluated as the RMS of the distribution of the fitted values.

\begin{figure*}
\includegraphics[width=0.80\textwidth]{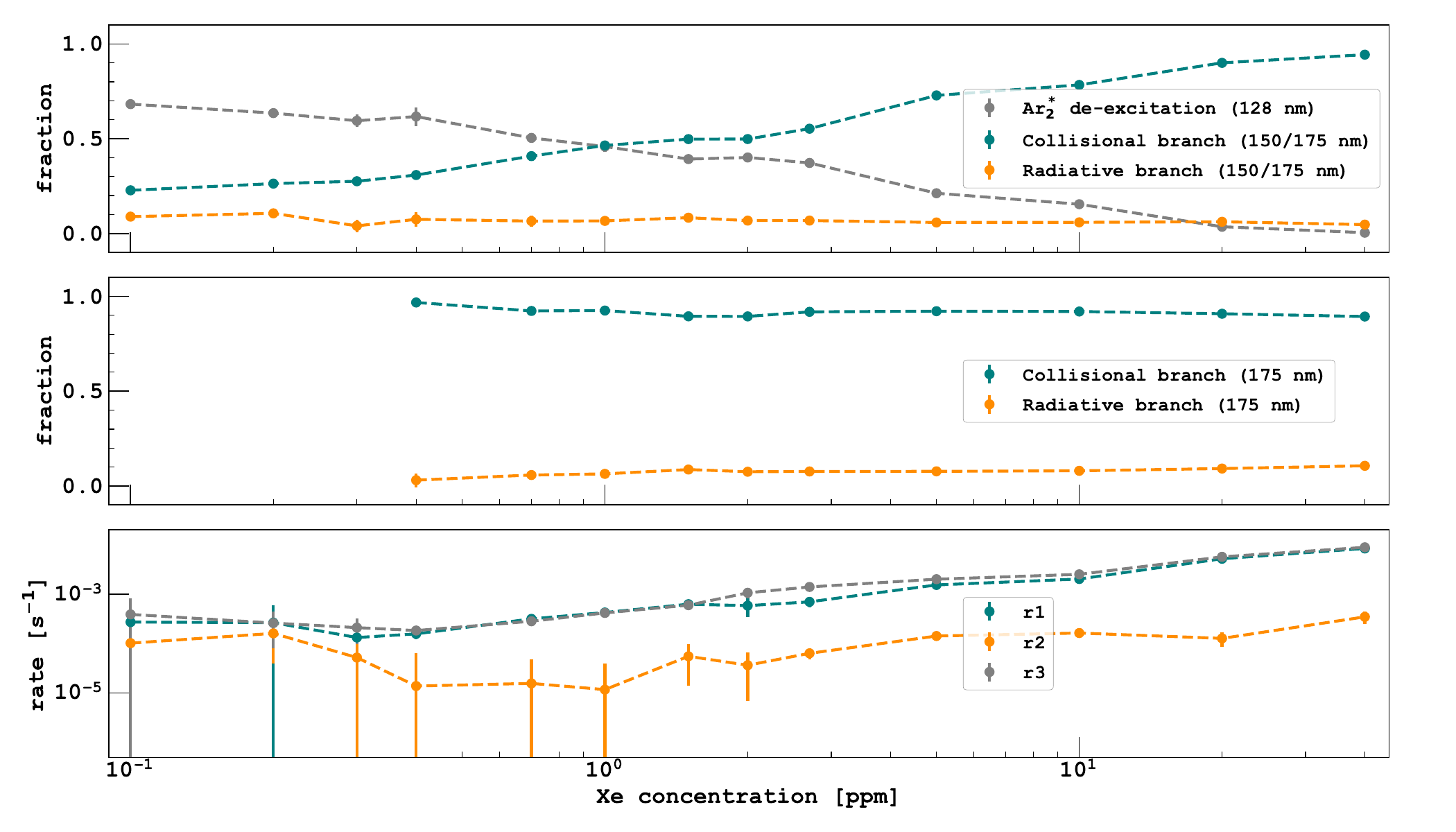}
  \caption{Fractions of direct LAr de-excitation, as well as collisional and radiative branches, are plotted for both WL-SiPMs (top) and WW-SiPMs (middle) as functions of xenon concentration. The bottom plot depicts the process rates versus xenon concentration. The values reported are obtained by averaging the fit results once stability is achieved following each xenon injection. Uncertainties in the Xe concentrations are estimated at 20\%.} 
  \label{fig:ppm}
\end{figure*}

Each run consists of 100,000 triggers. Averaged waveforms are obtained by selecting events where the total number of photoelectrons (pe) across all channels is between 300 and 700, a range selected to avoid channel saturation. To isolate electron recoil events (nuclear recoils from the $^{241}$AmBe source are discussed in the next section) we require that the fraction of prompt light detected within the first 200 ns is between 0.1 and 0.6. With these cuts enforced, each averaged waveform represents at least 5~000 triggers. Figure~\ref{fig:evolution}  shows the evolution of the amplitudes and process rates from eqs.~\ref{eq:wl} and \ref{eq:ww} as a function of time and the associated Xe injection phases. The data were subsequently reprocessed and expressed as a function of Xe concentration, as shown in figure~\ref{fig:ppm}, by averaging the results once their stabilization was achieved. This stabilization is interpreted as an indication of the complete diffusion of Xe within the LAr, ensuring a uniform distribution of the gas.  The  amplitudes are reported normalized to the sum of $A$, $B_1$, and $C_1$  and of $B_2$ and $C_2$ for WL and WW-SiPMs, respectively. 

As expected, the pure de-excitation component of Ar$_2^*$ decreases significantly with increasing Xe concentration, becoming negligible around 20 ppm when the collisional branch becomes dominant. In the WW-SiPM data, which is insensitive to the Ar$_2^*$ de-excitation, the amplitudes of the collisional and radiative branches remain nearly constant for Xe concentrations above 0.7 ppm, with the radiative component contributing approximately 8\%. At lower concentrations, the signal is too weak to produce stable results.  

The amplitude of the radiative component depends on the number of excited Ar ions and the EUV photon-Xe scattering length. The former is independent of Xe concentration, while the latter shows only a weak dependence, as the scattering length remains negligible compared to the detector size even at 0.1 ppm Xe doping. This combination explains the stability observed in figure~\ref{fig:ppm} for the radiative component amplitude across different Xe concentrations.

At the same time, the rates $r_1$, associated solely with the collisional branch, and $r_3$, linked to the ArXe$^*$ + Xe $\rightarrow$ Xe$_2^*$ + Ar process, both increase with increasing Xe concentration, reflecting the enhanced probability of energy transfer processes in xenon-doped liquid argon.  On the other hand, $r_2$, associated with the radiative branch, remains almost constant across different Xe concentrations. It is important to note that $r_2$ approximates a complex forest of emission lines with a single component, which are expected to last up to few seconds.  The short acquisition gate, limited to just 10~$\mu$s from the trigger, makes the fitting of this component quite unstable. Therefore, the results for $r_2$ should be considered as a first-order approximation.

%%%%%%%%%%%%%%%%%%%%%%%%%%%%%%%%%%%%%%%%%%%%%%%%%%%%%%%%%%%%%%%%%%
\section{The Xe-doped liquid argon light yield and pulse shape discrimination}
\label{sec:ly_psd}
%%%%%%%%%%%%%%%%%%%%%%%%%%%%%%%%%%%%%%%%%%%%%%%%%%%%%%%%%%%%%%%%%%

Xenon doping affects the LAr response in terms of both light yield (LY) and scintillation pulse shape discrimination (PSD) between nuclear and electronic recoils. To characterize its impact, the detector was exposed to   $^{60}$Co and $^{22}$Na $\gamma$-sources, as well as to the $^{241}$AmBe neutron emitter. %, and $^{133}$Ba

The overall LY observed in the detector is the sum of the contributions from photons emitted by Ar$_2^*$, Xe$_2^*$, and ArXe$^*$, whose relative fractions depend on the Xe concentration, weighted by the associated PDEs. 

Waveforms from each channel are integrated up to 10 $\mu$s from the trigger time and then normalized by the charge gain determined from laser calibrations. The contributions from the WL and WW channels are summed separately.

The number of photoelectron  is calculated separately for WW and WL sensors by summing the contributions from the associated channels. However, since one of the WW-SiPMs was not operational during data taking, the WW LY was scaled by a factor of 4/3 to account for the missing channel. This could introduce a bias in the ratio between the two LYs, as the two groups of channels do not have the same optical coverage and may experience different collection efficiencies depending on the interaction position within the chamber.

The  $^{60}$Co energy spectrum is simulated with a Geant4 Monte Carlo  and convoluted with a Gaussian response, with variance given by the sum of a linear and a quadratic term, to account for Poisson fluctuations and non-uniformity in light collection, respectively.

Data are fitted using this model through a binned likelihood approach while varying the Xe concentration. The resulting LY is reported for both the WW- and WL-SiPM channel groups and is normalized relative to the LY from WL-SiPMs in pure LAr. 

\begin{figure*}
    \includegraphics[width=0.8\textwidth]{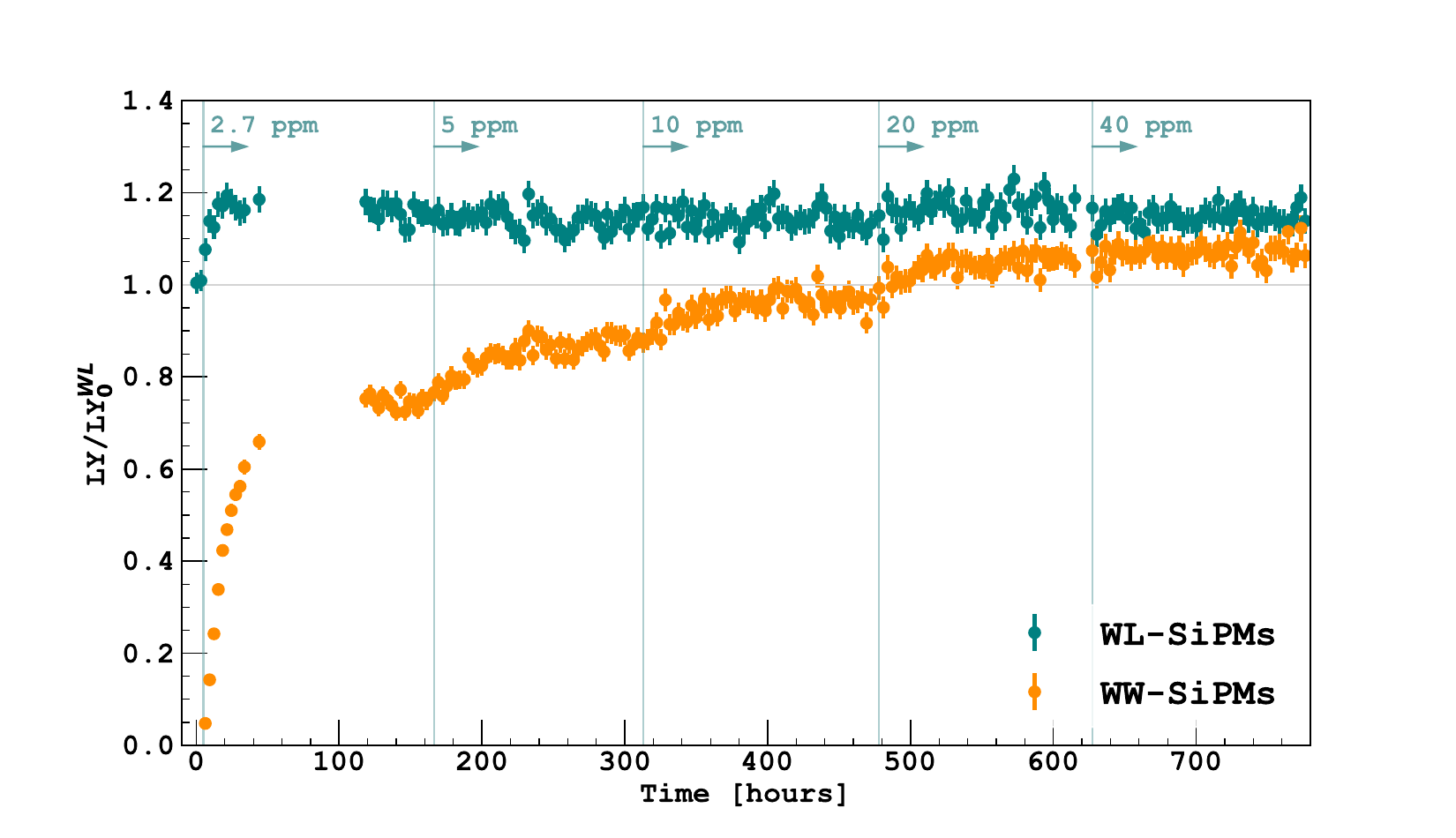}
    \caption{LYs  from the fits to the spectra of a $^{60}$Co source, as measured by WW and WL channels, shown as a function of Xe concentration. Values are normalized to the LY at 0 ppm from the WL SiPMs. Vertical lines correspond to Xe injections.   }
    \label{fig:ly_xe}
\end{figure*}

The LY of the WW-SiPMs, which are sensitive only to 175-nm photons, increases with the concentration of Xe, as shown in figure~\ref{fig:ly_xe}, and reaches a plateau at  $\sim$20 ppm. On the other side, the relative LY from WL-SiPMs, sensitive also to 128- and 150-nm photons, reaches a plateau already at few ppm, suggesting that Ar$_2^*$ de-excitations become sub-dominant with respect to the collisional ArXe branch, as also observed in the previous section. 

At 40 ppm, the WW LY corresponds to approximately 94\% of the WL LY, suggesting that the scintillation in this regime is largely dominated by Xe$^*_2$ de-excitations.  The overall WL LY increases by approximately 15\% at just 1 ppm. The increase is 30\% smaller in terms of photon yield, when accounting for the ratio of PDEs at 128 nm and 150/175 nm.

Similar increases have been observed in the literature~\cite{Vogl:2021rba, DUNE:2024dge}, although they are not directly comparable due to different detector geometries and optical coverages. Overall, the increase in LY is an interesting result for enhancing energy resolution in dark matter and neutrino experiments. However, LY is not the only relevant parameter; the PSD also plays a significant role, particularly in dark matter search experiments.

The PSD observable, $w$,  is here defined as the fraction of light collected within the first 300 ns. We characterized the $w$-distribution   by analyzing events  from all WL-SiPMs when exposing the detector to $^{22}$Na and $^{241}$AmBe sources.  The $^{22}$Na source  emits  $\gamma$-rays, inducing electronic recoils only, while the $^{241}$AmBe source produces also nuclear recoils, by emitting   both $\gamma$s  and neutrons. The $w$ distributions are produced by selecting  events in the  40-70 pe range. Such a range is  defined in order to maximize the number of nuclear recoil events and avoid saturation of the electronics. 

\begin{figure*}
 \includegraphics[width=.49\textwidth]{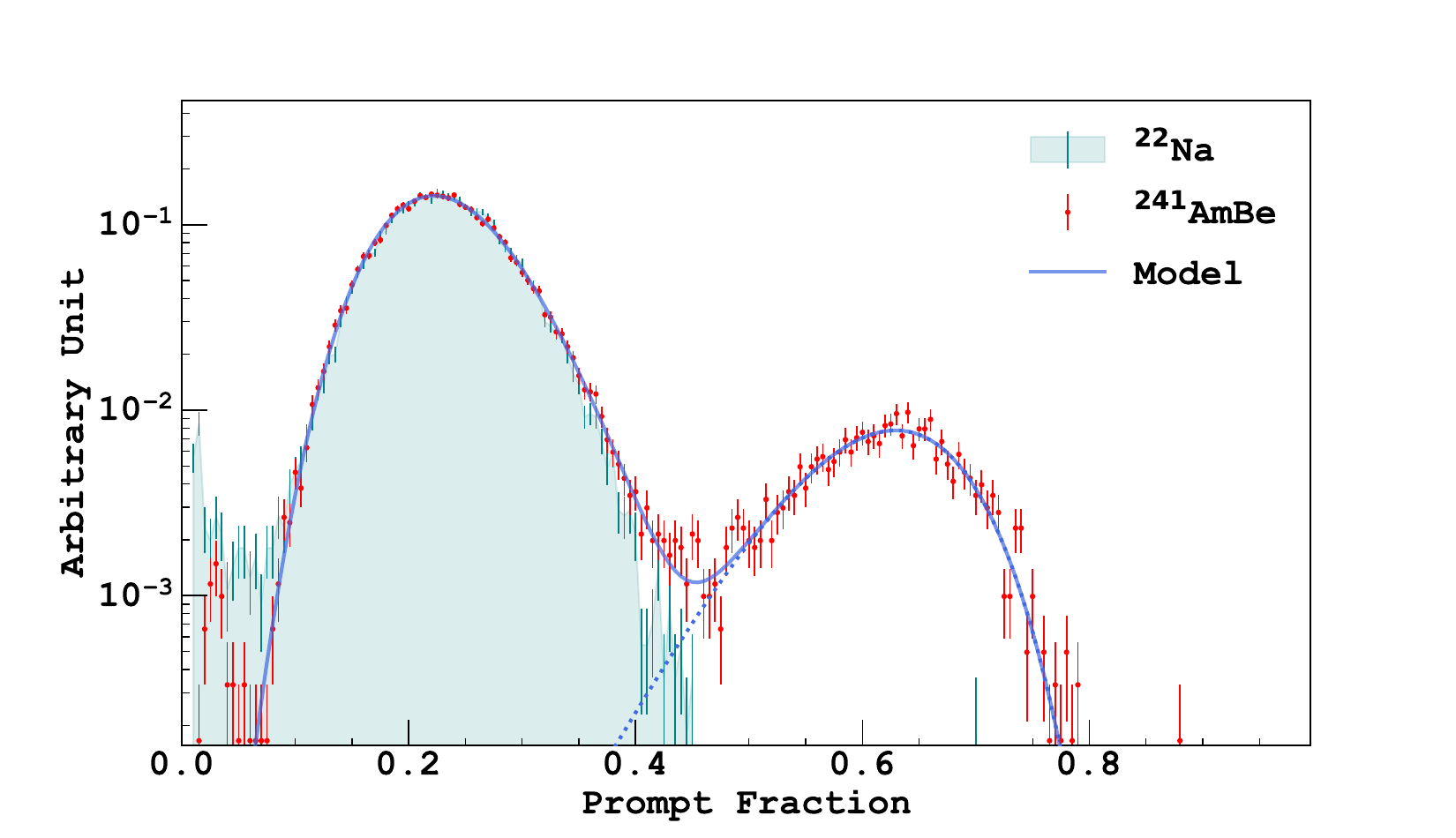}
  \includegraphics[width=.49\textwidth]{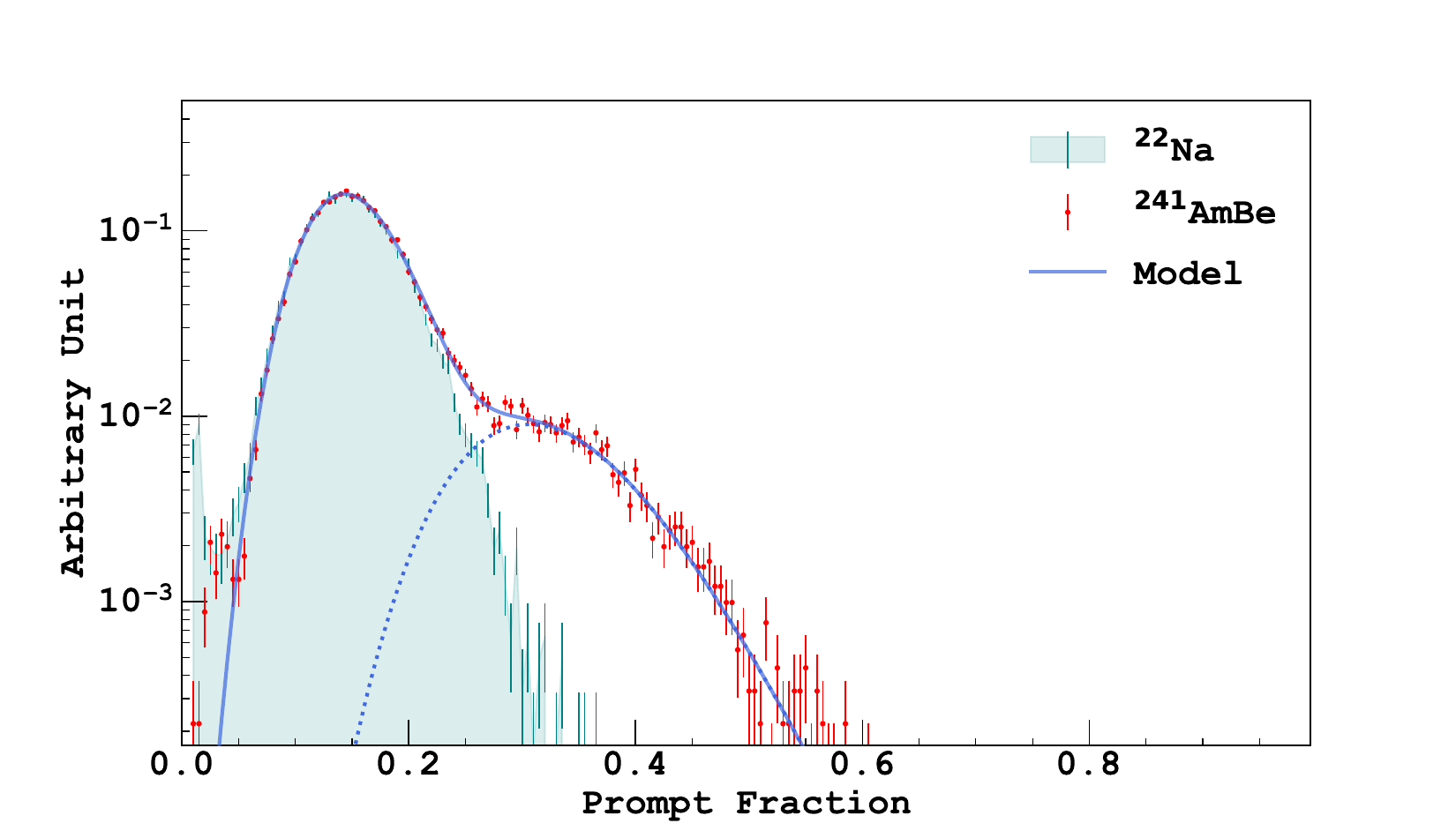}
 \caption{Examples of fits (red curves) of the fprompt distribution for samples acquired with 0.1 ppm (left) and 2.7 ppm (right) Xe-doping. The blue shaded area represents electronic recoils from the $^{22}$Na $\gamma$-source, while the red dots from the $^{241}$AmBe source include both electronic recoils and nuclear recoils from neutron interactions. }
  \label{fig:arxe_psd}
\end{figure*}

Data are fitted with the model from ref.~\cite{Agnes:2021cqa}, where  $w$ is distributed as a gaussian with variance
\begin{equation}
\sigma^2(w) = (1 - w)^2  k_p^2 + w^2  k_l^2  + 2  w\, (1-w)\, k_p \, k_l
\end{equation}
\noindent with $k_p$ and $k_l$ free parameters in the fit describing the $w$ fluctuations  (see ref.~\cite{Agnes:2021cqa} for more details). 

The fitting procedure is done in two steps. First, only the electronic recoil distribution is fitted using the $^{22}\text{Na}$ data. The parameters are then constrained in the electronic recoil component when fitting the $^{241}\text{AmBe}$ data, which also includes contributions from nuclear recoils. Examples of fits at 0.1 and 2.7 ppm Xe concentrations are shown in figure~\ref{fig:arxe_psd}.

\begin{figure}
\includegraphics[width=.49\textwidth]{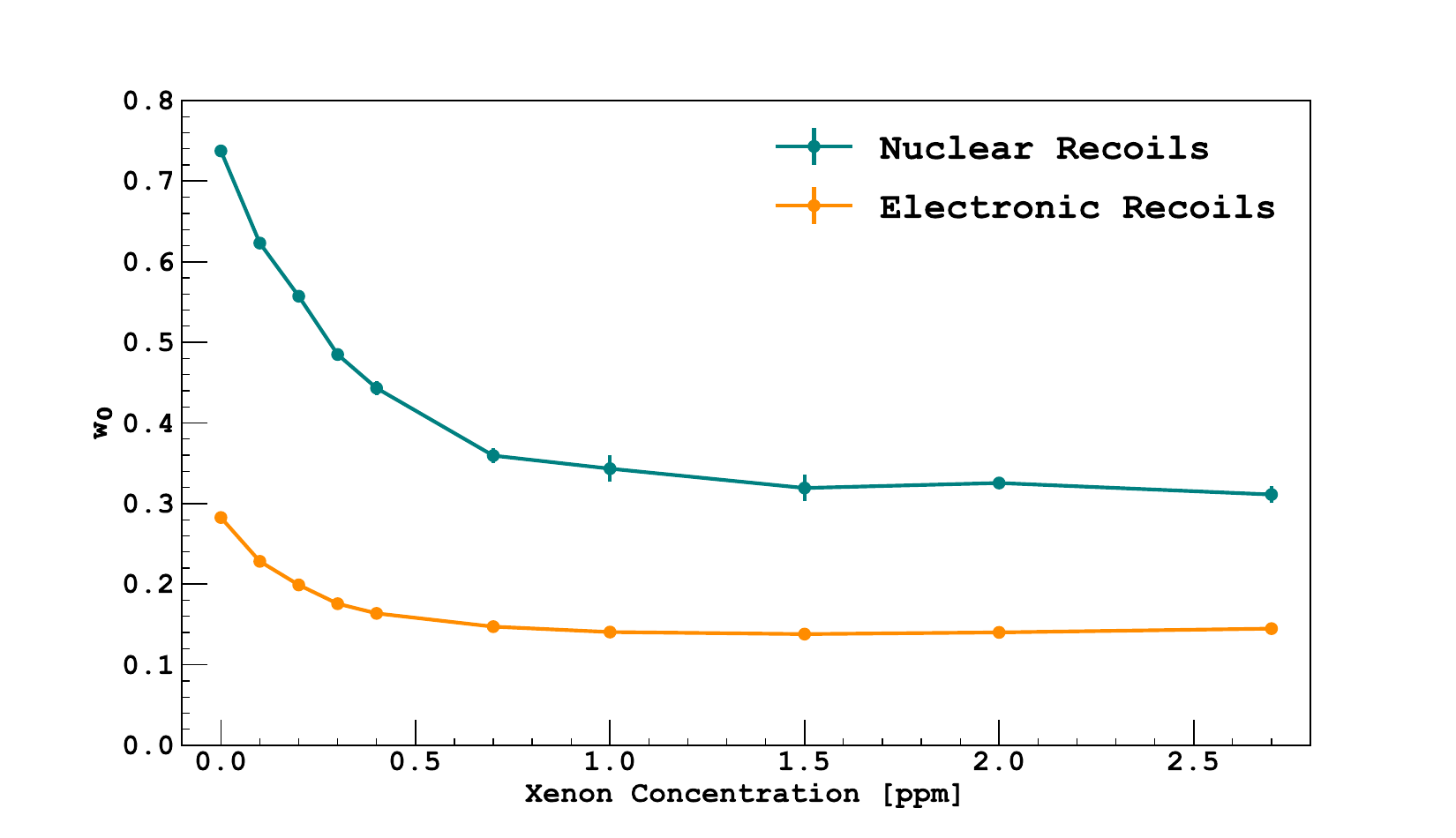}
  \caption{Comparison between centroids of  ER and NR scintillation prompt ($<$200~ns) fraction distributions, selected in the  [40,100] pe range, as a function of  Xe concentration.}
  \label{fig:arxe_w0}
\end{figure}

The photoelectron range of the analyzed datasets   is lower than that used by dark matter experiments~\cite{DarkSide:2014llq, DEAP:2021axq} to discriminate between electronic and nuclear recoils. As a result, the discrimination power we could here extract  in terms of nuclear recoil acceptance and electronic recoil leakage is not significant, being limited by the low photoelectron statistics. Instead, we use the fitted centroids ($w_0$) of the $w$-distributions  as a figure of merit, since the closer the  electronic and nuclear centroids, the lower the discrimination power.  The fitted centroids are shown as a function of the Xe concentration in figure~\ref{fig:arxe_w0}, up to 2.7 ppm, the highest concentration tested in the $^{241}\text{AmBe}$ data campaign. Notably, the PSD degrades as the Xe concentration increases up to 1 ppm, but then stabilizes beyond this range. This suggests that Xe-doped LAr can still retain some PSD power even at Xe concentrations of a few ppm. 

We also investigated whether the ratio of collected light between WW and WL SiPMs varied with the nature of the recoil, but no significant differences were observed.

In summary, Xe-doping leads to a significant enhancement of the LY of LAr, even at concentrations as low as a few ppm. However, this improvement comes at the expense of partially degrading its PSD capability. The limited light collection of the detector prevents further advancements, highlighting the need for a chamber with higher photon collection efficiency to optimize results and accurately establish the PSD power in the 0.1-10 ppm  Xe concentration range.

%%%%%%%%%%%%%%%%%%%%%%%%%%%%%%%%%%%%%%%%%%%%%%%%%%%%%%%%%%%%%%%%%%
\section{Impact of the EUV hypothesis on spurious electrons}
\label{sec:se}
%%%%%%%%%%%%%%%%%%%%%%%%%%%%%%%%%%%%%%%%%%%%%%%%%%%%%%%%%%%%%%%%%%
The lowest energy threshold achievable in noble liquid dual-phase TPCs is on the order of a few tens of eV, corresponding to the electroluminescence pulse induced in the gas phase by a single electron. However, this region is affected by a significant rate of so-called spurious electrons (SEs), which are electroluminescence pulses equivalent to one or a few electrons, often time-correlated with high-energy events in the TPC~\cite{XENON100:2013wdu,LUX:2020vbj}. The long correlation time, extending up to several milliseconds, poses a major challenge, particularly in the search for light dark matter particles~\cite{DarkSide-20k:2024yfq, GlobalArgonDarkMatter:2022ppc}. The mechanisms underlying SE production remain uncertain, with two main hypotheses suggesting that SEs either originate from ionization electrons trapped by impurities and released after a delay or result from delayed electron extraction at the liquid-gas interface~\cite{XENON:2021qze}.  Additional studies hint at fluorescence effects in LXe that might also play a role through impurity photoionization~\cite{Sorensen:2017kpl}.

EUV photons may  represent a solid alternative hypothesis at the origin of SEs.  Indeed, EUV from argon atomic states have sufficient energy to ionize trace impurities usually present in LAr, such as O$_2$ (12.1~eV), CH$_4$ (12.6~eV), H$_2$O (12.6~eV), CO$_2$ (13.8~eV),  Kr (14.0~eV) and CO (14.0~eV), leading to the emission of free electrons. A similar mechanism is expected in LXe, where EUV photons from xenon atomic states could also ionize impurities~\cite{Merabet2005,Merabet_2007, ALI2017122,Osin_2012}. %Photo-ionization of impurities in LXe was also hypothized in ref.~\cite{Sorensen:2017kpl} from fluorescence photons.  

%A similar mechanism can be hypothesized in liquid xenon as well~\cite{Merabet2005,Merabet_2007, ALI2017122,Osin_2012}. Photo-ionization of impurities from Xe fluorescence  photons was also investigated in ref.~\cite{Sorensen:2017kpl}. 

The EUV hypothesis could also account for the observed energy and time correlations with preceding events~\cite{XENON:2021qze}.  Higher energy events increase the probability of producing excited atomic states, leading to a higher rate of EUV photon emission and, consequently, of impurity ionization. In addition, the observed time delays are consistent with the characteristic lifetimes of the atomic metastable states. 

The EUV hypothesis, however,  does not account for the multiplicity of SEs, which has been observed up to  a few electrons in experiments like DarkSide-50~\cite{DarkSide-50:2022qzh} and XENON1T~\cite{XENON:2019gfn,XENON:2021qze}. It is indeed unlikely that multiple EUV photons are emitted simultaneously from different metastable states, leading to signal pile-ups. However, multiple SEs can still be explained by other known mechanisms, such as the photoelectric effect on metallic surfaces, like the grid used to shape the field for extracting electrons from the liquid to gas phase.

In conclusion, EUV photons from metastable atomic states represent a compelling hypothesis for the origin of SEs, albeit yet to be proven.  

%%%%%%%%%%%%%%%%%%%%%%%%%%%%%%%%%%%%%%%%%%%%%%%%%%%%%%%%%%%%%%%%%%
\section{Conclusions}

This study advances our understanding of scintillation dynamics in liquid argon, both pure and xenon-doped. Using a novel dual-SiPM setup, with and without quartz windows, we suggest the presence of a radiative component induced by EUV photons from metastable Ar$^*$/Ar$^{q+}$ excited states. We developed a comprehensive model incorporating both collisional and radiative processes, showing that the EUV-induced component persists across xenon concentrations. Additionally, we observed up to a 15\% enhancement in the light yield, and concurrently a degradation in the scintillation pulse shape discrimination as the Xe concentration increased. These findings have important implications for detector design in neutrino physics and dark matter searches.  For example, an ARGO-like solar neutrino experiment relying on elastic scattering off electrons~\cite{Franco:2015pha} or a $^{136}$Xe neutrinoless double beta decay search~\cite{PhysRevD.106.092002}, both using Xe-doped LAr targets, could benefit from the enhanced light yield, resulting in improved energy resolution. Additionally, they could take advantage of the residual PSD rejection power, still exceeding $10^{3}$ at the MeV scale as estimated with Monte Carlo simulations assuming a light yield of 8 pe/keV,  highly effective in suppressing alpha background. In a broader context, the scintillation model described in this work could also benefit DUNE low-energy astroparticle program, which will rely on ppm-level Xe-doped LAr target.

The impact of this work is  significant also for noble liquid experiments searching for light dark matter. Our research suggests that EUV photons may ionize residual impurity molecules, offering a potential explanation for the long-standing issue of spurious electrons in pure liquid argon targets (and possibly in liquid xenon as well), which represent the dominant background in the lowest energy range. At the same time, late photons from the tails of the scintillation pulses are difficult to associate with the pulse itself and could be misinterpreted as a transient increase in the dark count rate. These insights could inform strategies for background reduction in future detectors, potentially enhancing their sensitivity. Future work should focus on experimentally verifying the EUV-impurity interaction and exploring methods to mitigate its effects.
%%%%%%%%%%%%%%%%%%%%%%%%%%%%%%%%%%%%%%%%%%%%%%%%%%%%%%%%%%%%%%%%%%

\acknowledgments

We thank Benoît Gervais (CIMAP/GANIL, Caen), Paul-Antoine Darius Hervieux (IPCMS, Strasbourg), J\"orn Schwandt (Universit\"at Hamburg) for their valuable suggestions. We acknowledge the financial support from the ANR \href{https://x-art.in2p3.fr/}{X-ArT}  (Grants N. ANR-22-CE31-0021), IN2P3–COPIN (No. 20-152),  NCN Poland (2021/42/E/ST2/00331),  IRAP AstroCeNT (MAB/2018/7) funded by FNP from ERDF, and the Research Corporation for Science Advancement.

\bibliography{biblio}
\end{document}